\documentclass[10pt]{article}
\usepackage{OSAconf}
\usepackage[latin1]{inputenc}
\usepackage[T1]{fontenc}
\usepackage[centertags]{amsmath}
\usepackage{amstext}
\usepackage{graphicx}
\usepackage[footnotesize, normal]{caption2}
\usepackage[letterpaper,width=6.5in,height=9in,headheight=0in,headsep=0in,footskip=0in,top=1in]{geometry}
\usepackage{units}
\usepackage{siunitx}
\usepackage{soul}
\usepackage{mathptmx}

\usepackage{xspace}

\begin{document}

\title{Automatic Wavelength Tuning of Series-Coupled Vernier Racetrack Resonators on SOI}
\author{Hasitha Jayatilleka, Robert Boeck, Kyle Murray, Jonas Flueckiger, Lukas Chrostowski,
Nicolas A. F. Jaeger, and Sudip Shekhar}
\address{Department of Electrical and Computer Engineering, University of British Columbia, \\
2332 Main Mall, Vancouver, BC V6T 1Z4, Canada.}
\email{hasitha@ece.ubc.ca}

\vspace*{-0.1in}
\begin{abstract}
Using in-resonator photoconductive heaters to both sense and control the intra-cavity light intensity of microring resonators, automatic tuning of a silicon-on-insulator two-ring Vernier filter is demonstrated across the entire C-band. 
\end{abstract}
\\
\parindent.2in
\vspace*{-0.1in}
\section{Introduction\\}
By coupling resonators that have different optical path lengths, the Vernier effect can be used to increase the free-spectral-ranges (FSRs) and wavelength tuning ranges achievable in microring resonator-based devices [1]. Widely tunable lasers [2], wavelength selective switches [3], and reconfigurable filters [4, 5], capable of meeting numerous telecom-grade filter specifications [1], have been demonstrated on silicon photonics platforms using the Vernier effect. Tunable Vernier devices allow one to adjust the performance of the devices to address variations in fabrication, operating temperatures, and wavelength of the input laser, as well as to completely reconfigure the devices to operate at various channel wavelengths [1, 5]. Therefore, the ability to automatically tune these devices is an essential requirement for their practical deployment.

Recently, we showed that automatic wavelength tuning and stabilization of silicon microring-based filters can be achieved using in-resonator photoconductive heaters (IRPHs) to both sense and control the light intensity inside the resonators [6]. The IRPHs are \textit{n}-doped waveguide sections that can be integrated into each microring in a silicon-on-insulator (SOI) filter. The photodetection in IRPHs occurs due to defect-state-absorption [7]. Nevertheless, IRPHs show high responsivities without requiring dedicated defect implantations [6]. The \textit{n}-doped waveguides also act as resistive heaters, allowing for thermo-optic tuning of the microring resonators.

In this work, we show how a Vernier filter fabricated with two racetrack resonators with incorporated IRPHs in a series-coupled configuration [Fig. 1(a)]  can be automatically tuned to center its response to any input laser wavelength. The automatic tuning is achieved using simple maximum-search algorithms, which are used to find the heater voltage settings that maximize the light intensity in each of the resonators. The light intensity in a resonator is determined by measuring the photocurrent generated by its IRPH. As compared to previous work [7-9], the automatic tuning described here is achieved without requiring dedicated defect implantations, additional material depositions, dedicated photodetectors, thermal sensors, or optical power taps. In this work, we present the first demonstration of automatic wavelength tuning of a Vernier ring resonator filter.

\section{Device and tuning method\\}
Figure 1(a) shows a microscope image of a SOI two-ring Vernier filter which was fabricated at the A*STAR IME foundry. The IRPHs, formed by \textit{n}-doping the silicon rib waveguides (rib height = \SI{220}{nm}, slab height = \SI{90}{nm}), are shown as resistors in the circuit. The doping was achieved by ion-implantation. The two racetrack resonators, labeled ``Ring 1'' and ``Ring 2'' had lengths $L_1$ = \SI{66.83}{\um} and $L_2$ = \SI{83.54}{\um}, respectively. 

\begin{figure}[b]
\vspace*{-0.2in}
  \centering
  \includegraphics[height=1.45in]{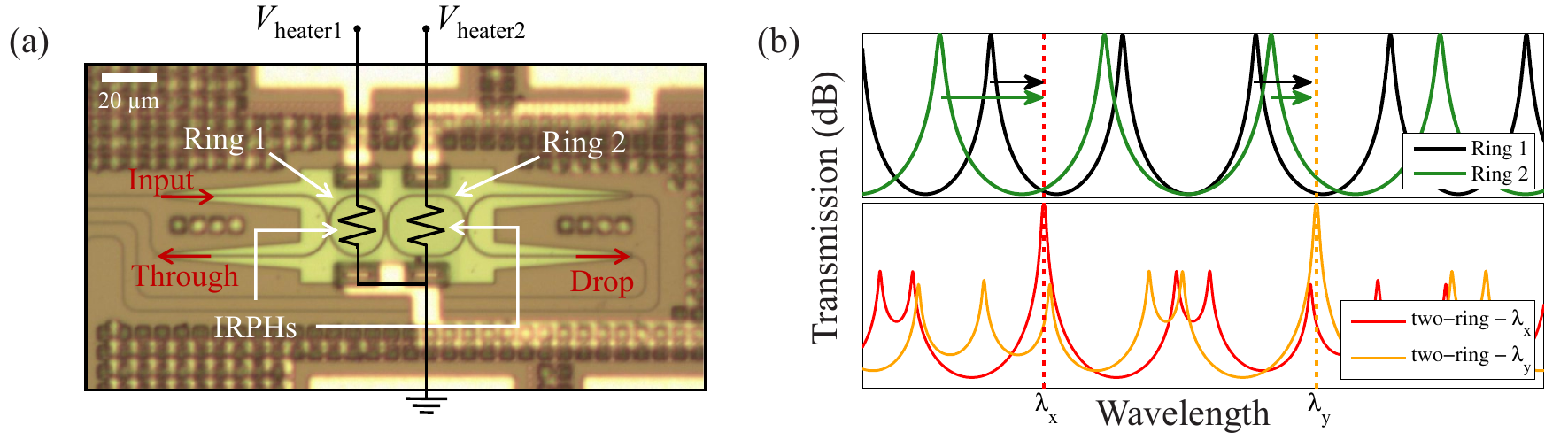}
  \caption{(a) A microscope picture of a fabricated two-ring Vernier filter. (b) Illustration of tuning the nearest shorter-wavelength resonances of Ring 1 and Ring 2 to obtain the overall two-ring responses at either ${\lambda_\text{x}}$ or ${\lambda_\text{y}}$.}
  \label{fig1}
  \vspace*{-0.1in}
\end{figure}

Figure 1(b) illustrates how tuning the nearest shorter-wavelength resonance of each resonator to the input laser's wavelength at ${\lambda_\text{x}}$ results in the overall response being centered at ${\lambda_\text{x}}$. This is achieved by first tuning Ring 1, which is coupled to the input waveguide, and setting its heater voltage $V_\text{heater1}$ such that $I_\text{PD1}$, the photocurrent measured by the IRPH in Ring 1, is maximized. Next, Ring 2 is tuned to maximize photocurrent of its IRPH, $I_\text{PD2}$. This ensures that both rings are resonant at ${\lambda_\text{x}}$, as the maximum photocurrents correspond to the maximum light intensities in the resonators. As the heater voltages are set by sequentially finding the first maxima of $I_\text{PD1}$ and $I_\text{PD2}$, this method also ensures that the tuning power required by each ring is at most the tuning power required to shift its resonance by about one FSR. For example, as illustrated in Fig. 1(b), when the input laser is at either ${\lambda_\text{x}}$ or ${\lambda_\text{y}}$, the nearest un-tuned shorter-wavelength resonance of each ring is automatically selected and tuned to the appropriate wavelength. 

The application of the above tuning method to a fabricated device is demonstrated in Fig. 2. Measured $I_\text{PD1}$ and $I_\text{PD2}$, as the rings are tuned by sweeping $V_\text{heater1}$ and $V_\text{heater2}$, are shown in Figs. 2(a) and 2(b), respectively. At the end of each sweep, $V_\text{heater1,2}$ was set to the value that maximized $I_\text{PD1,2}$. The IRPHs carry the heater currents $I_\text{heater1,2}$ together with $I_\text{PD1,2}$. Therefore, in order to measure $I_\text{PD1}$ and $I_\text{PD2}$, a 2-D calibration step was performed before the measurement by turning the laser off and measuring $I_\text{heater1,2}$ as functions of both $V_\text{heater1}$ and $V_\text{heater2}$. At each combination of $V_\text{heater1}$ and $V_\text{heater2}$, $I_\text{PD1,2}$ was calculated by subtracting $I_\text{heater1,2}$ from the total measured current. This 2-D calibration was required to account for the ``crosstalk'' current between the two IRPHs due to the finite resistance of the un-doped silicon slab region that separated the two IRPHs. Figure 2(c) shows the improvement of the through- and drop-port spectra after completing the tuning steps. As Ring 2 is tuned after tuning Ring 1, the heat from Ring 2 can cause Ring 1 to be slightly detuned from the initially found resonance condition. In order to counteract the effects of this thermal crosstalk and optimize the drop-port response, a maximum-search algorithm was used. This algorithm [6] continuously stepped the electrical power supplied to the heaters $P_\text{heater1}$ and $P_\text{heater2}$ to maximize $I_\text{PD2}$. The filter spectra after tuning and after applying the optimization step is shown in Fig. 2(d). By manually adjusting the heater voltages, it was confirmed that the drop-port transmission was maximized after completing the automatic tuning and optimization steps. Figure 2(e) shows the filter spectra from \SI{1500}{nm} to \SI{1570}{nm}, both as-fabricated and after optimization. The electrical power required to tune each ring through one FSR was \SI{38}{mW}. All of the transmission spectra shown in this paper are normalized to the input/output grating coupler spectra extracted from the off-resonance through-port transmission of the device.

\begin{figure}[h]
  \vspace*{-0.1in}
  \centering
  \includegraphics[height=2.8in]{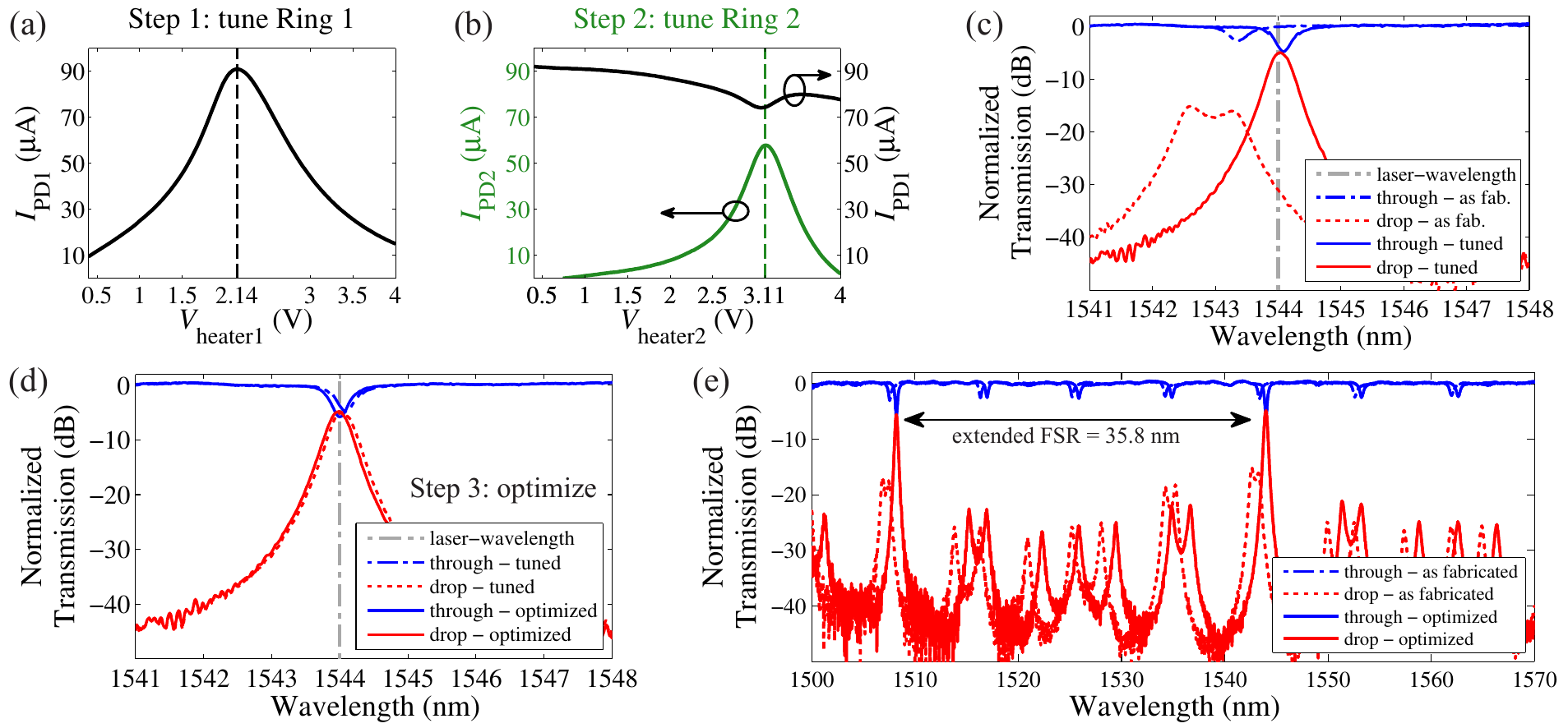}
  \caption{Measured (a) $I_\text{PD1}$ vs. $V_\text{heater1}$, when $V_\text{heater2}$ = 0 V, and (b) $I_\text{PD2}$ and $I_\text{PD1}$ vs. $V_\text{heater2}$, after setting $V_\text{heater1}$ = 2.14 V. Measured through- and drop-port spectra of the filter (c) as-fabricated and after tuning Ring 1 and Ring 2, (d) after tuning and after optimization, and (e) as-fabricated and after optimization.}
  \label{fig1}
  \vspace*{-0.1in}
\end{figure}

\section{Automatic tuning across the C-band\\}
In this section, we show the automatic tuning of the Vernier filter to all of the ITU channels from \SI{1528.77}{nm} to \SI{1565.50}{nm} with a \SI{200}{GHz} channel spacing, which includes the entire C-band. The filter was tuned to each wavelength by first setting the wavelength of the input laser to the channel wavelength and then applying the above described tuning and optimization steps. The drop-port spectra recorded for each channel after the tuning and optimization steps are shown in Fig. 3(a) and 3(b), respectively. Measured $P_\text{heater1}$ and $P_\text{heater2}$ after the tuning and the optimization steps are shown in Fig. 3(c). The discrete jumps in $P_\text{heater1,2}$, observed when $P_\text{heater1,2}$ reaches the full FSR tuning power of a single resonator, occur as a result of tuning the nearest shorter wavelength un-tuned resonance of either ring to the input laser's wavelength. Hence, the amount of tuning power needed to achieve the required wavelength shift resets every FSR of each ring. This implies that any wavelength within the \SI{35.8}{nm} extended FSR of the Vernier filter can be addressed with a maximum of 2$\times$\SI{38}{mW} of tuning power.
 
\begin{figure}[h]
  \vspace*{-0.1in}
  \centering
  \includegraphics[height=3.7in]{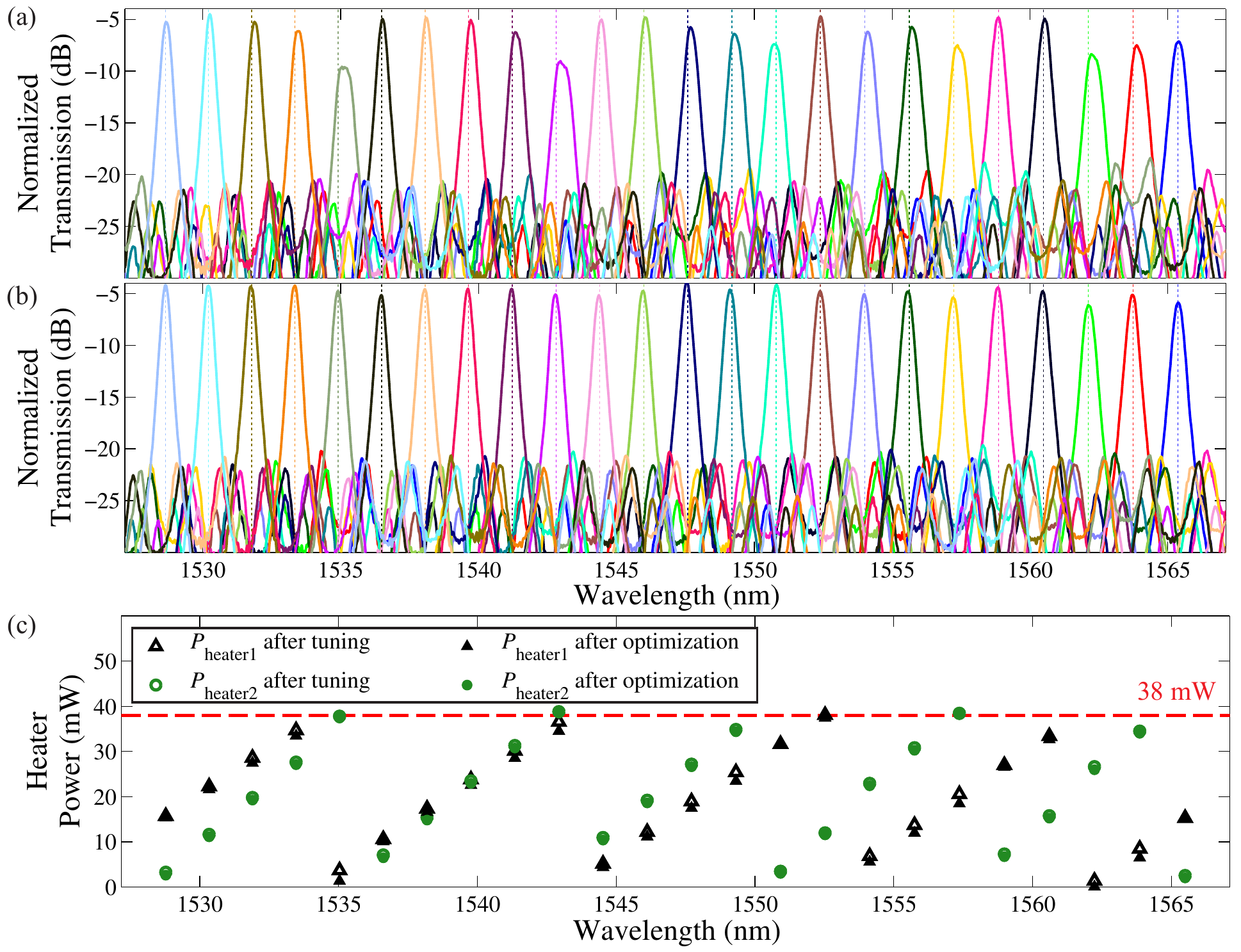}
  \caption{Overlaid measured drop-port spectra (a) after the tuning (steps 1 and 2), and (b) after the optimization (step 3) - the dotted lines indicate the channel wavelengths. (c) Measured electrical power supplied to the IRPHs of Ring 1 and Ring 2.}
  \label{fig1}
  \vspace*{-0.1in}
\end{figure}

As seen from Figs. 3(a) and 3(c), the thermal crosstalk during the tuning steps only caused slight detunings. Hence, only a few iterations of the optimization algorithm were required to maximize the drop-port transmission. As demonstrated in [6], for a microring filter with two identical rings, stabilization of the Vernier filter in the presence of ambient temperature fluctuations can be achieved by the continuous implementation of the maximum search algorithm that was used for the optimization of the drop-port power. In this work, all of the tuning steps were computer implemented using a Keithley 2602 source measurement unit. Nevertheless, the method can be readily implemented using dedicated electronic circuits for high-speed tuning and stabilization.
\section{Conclusion\\}
We have demonstrated automatic tuning of a two-ring Vernier filter across the C-band using IRPHs. As the resonance conditions of individual resonators can be simultaneously monitored and controlled using the IRPHs, the methods shown here can be readily extended for automatic wavelength tuning and stabilization of various microring based devices or systems with many microring resonators, such as filters, sensors, modulators, and lasers. 
\section{Acknowledgements\\}
The authors thank Michael Caverley and Miguel Guill\'{e}n-Torres for their help, the Natural Sciences and Engineering Research Council of Canada for financial support, CMC microsystems for providing access to fabrication and design tools, and the OpSIS MPW program for fabrication.

\end{document}